\documentclass[sigconf]{acmart}
\AtBeginDocument{%
  \providecommand\BibTeX{{%
    \normalfont B\kern-0.5em{\scshape i\kern-0.25em b}\kern-0.8em\TeX}}}

\setcopyright{acmcopyright}
\copyrightyear{2021}
\acmYear{2021}
\acmDOI{10.1145/3502223.3502247}

\acmConference[IJCKG'21]{The 10th International Joint Conference on Knowledge Graphs}{December 6--8, 2021}{Virtual Event, Thailand}
\acmBooktitle{The 10th International Joint Conference on Knowledge Graphs (IJCKG'21), December 6--8, 2021, Virtual Event, Thailand}
\acmPrice{15.00}
\acmISBN{978-1-4503-9565-6/21/12}

\begin{document}

\title{Knowledge Graph Curation: A Practical Framework}

\author{Elwin Huaman}
\email{elwin.huaman@sti2.at}
\orcid{0000-0002-2410-4977}
\affiliation{%
  \institution{Semantic Technology Institute (STI) Innsbruck, \\
  University of Innsbruck}
  \streetaddress{Technikerstrasse, 21a}
  \city{Innsbruck}
  \country{Austria}
  \postcode{6020}
}

\author{Dieter Fensel}
\email{dieter.fensel@sti2.at}
\affiliation{%
  \institution{Semantic Technology Institute (STI) Innsbruck, \\
  University of Innsbruck}
  \streetaddress{Technikerstrasse, 21a}
  \city{Innsbruck}
  \country{Austria}
  \postcode{6020}
}


\begin{abstract}
Knowledge Graphs (KGs) have shown to be very important for applications such as personal assistants, question-answering systems, and search engines. Therefore, it is crucial to ensure their high quality. However, KGs inevitably contain errors, duplicates, and missing values, which may hinder their adoption and utility in business applications, as they are not curated, e.g., low-quality KGs produce low-quality applications that are built on top of them.
In this vision paper, we propose a practical knowledge graph curation framework for improving the quality of KGs. First, we define a set of quality metrics for assessing the status of KGs, Second, we describe the verification and validation of KGs as cleaning tasks, Third, we present duplicate detection and knowledge fusion strategies for enriching KGs.
Furthermore, we give insights and directions toward a better architecture for curating KGs.
\end{abstract}



\maketitle

\section{Introduction}
\label{sec:introduction}
Knowledge graph curation (aka knowledge graph refinement~\cite{FenselSAHKPTUW20}) is the process of improving the quality of knowledge graphs (KGs). In this context, knowledge assessment, cleaning, and enrichment are critical tasks to provide reliable, correct, and complete knowledge. ``Knowledge Graphs are very large semantic nets that integrate various and heterogeneous information sources to represent knowledge about certain domains of discourse''~\cite{FenselSAHKPTUW20}.

Over the last decade, creating and especially maintaining large KGs have gained attention (e.g. Amazon Product Knowledge Graph \cite{Dong2019}, Bing Knowledge Graph, eBay's Product Knowledge Graph, Google’s Knowledge Graph~\cite{NoyGJNPT2019}). KGs provide structured data for customers' applications such as search engines, personal assistants, and question answering systems. However, KGs inevitably have inconsistencies, such as duplicates, wrong assertions, missing values, and more. The presence of such issues may compromise the outcome of business intelligence applications. Hence, it is crucial and necessary to explore efficient and effective semi-automatic methods and tools for tackling the curation of KGs. In other words, we need a curation framework for KGs that can well balance between ensuring correctness and completeness of knowledge graphs.

To face this challenge, we proposed a practical framework for improving the quality of KGs. Our approach involves 
(1) assessing the status of KGs based on quality dimensions and metrics, 
(2) detecting and correcting wrong assertions, and 
(3) enriching the KGs by adding new statements. Furthermore, we discuss our findings.

There have been approaches proposed to curate KGs. In this paper, we review methods and tools for knowledge curation. We found out that most of them focus on one specific task, either assessing the quality, detecting wrong assertions, or correcting those wrong assertions. However, curation of KGs usually implies a trade-off between correctness and completeness, which is tackled and assessed differently in each knowledge graph~\cite{FenselSAHKPTUW20,Paulheim17,ZaveriRMPLA16}. Therefore, we propose a practical knowledge curation framework, which is based on a process model for knowledge graph generation proposed by~\cite{FenselSAHKPTUW20}, that tackles the assessment, correctness, and completeness of KGs. In addition, we take into account the user perspective to define a degree of importance (i.e., weights) when curating KGs, e.g., the degree of importance determines KGs' utility in specific application scenarios.

This paper is structured as follows. Section~\ref{sec:framework} presents a practical framework for curation of KGs. In Section~\ref{sec:findings} we list some interesting findings. Finally, we conclude in Section~\ref{sec:conclusion}, by summarizing the conclusions and future work plans. 

\section{Knowledge Graph Curation Framework}
\label{sec:framework}
While a lot of effort is being invested in the deployment of KGs, new issues arise, such as the verification and validation of knowledge (i.e., knowledge cleaning tasks), the increasing coverage of KGs (i.e., knowledge completeness task), and the quality assurance (i.e., knowledge assessment task). These tasks are of utmost importance as the KGs grow to billions of statements~\cite{NoyGJNPT2019}.

\begin{figure*}
    \centering
    \includegraphics[width=\textwidth]{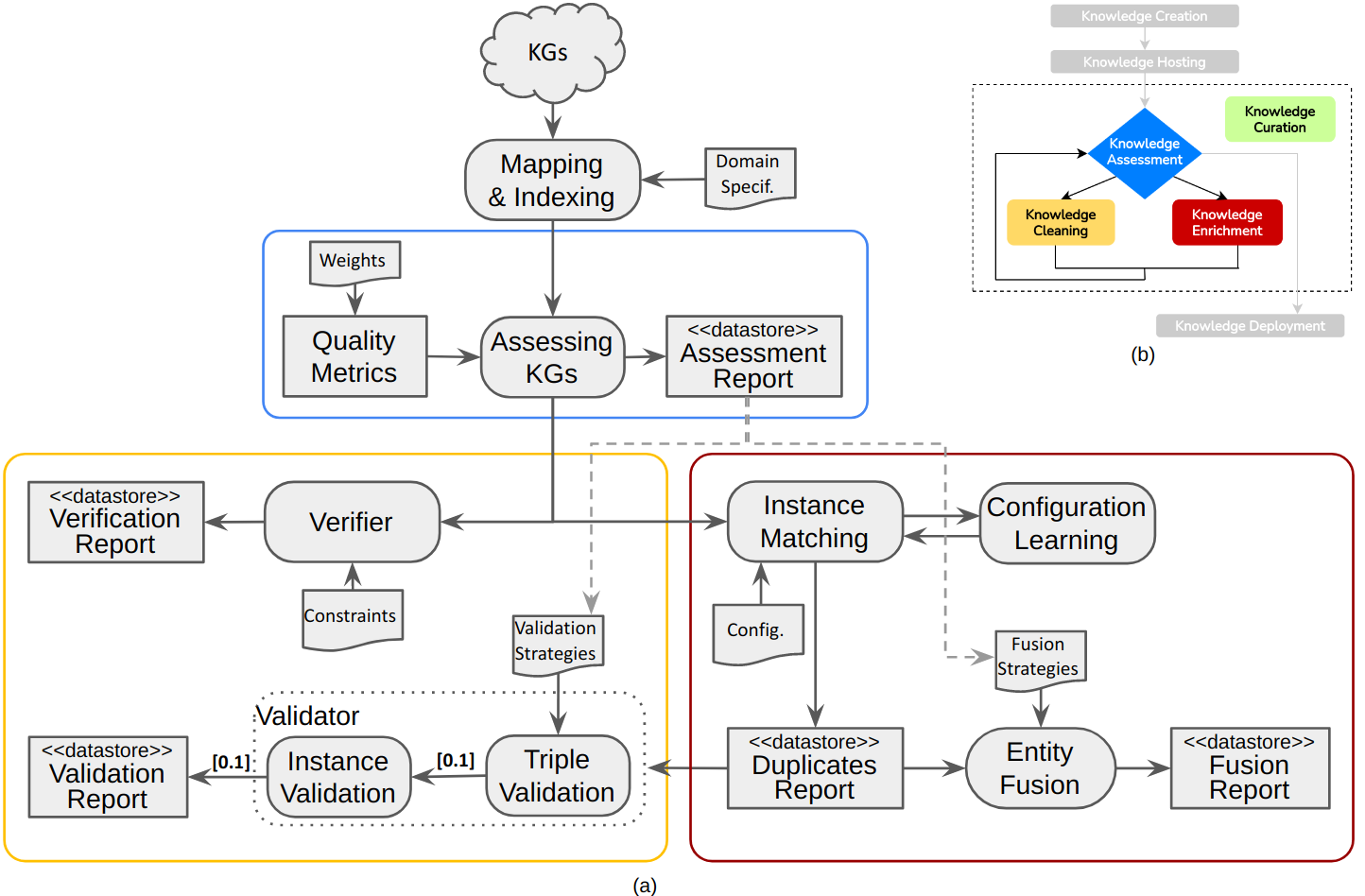}
    \caption{(a) The proposed knowledge graph curation framework, which is based on (b) The knowledge graph lifecycle~\cite{FenselSAHKPTUW20}.}
    \label{fig:knowledge-curation-framework}
\end{figure*}

The Knowledge Curation Framework follows the workflow described in Figure~\ref{fig:knowledge-curation-framework}. Before start, we assumed that a KG has been created and hosted in advance. 
First, the workflow starts with the assessment component that receives as input a mapped and indexed knowledge graph (KG), afterwards the assessing process starts triggered by the definition of the quality metrics and weights of importance for each of the quality metrics, later the outcome of the assessment is stored, so the information can be used in other components.
Second, at this point, either the cleaning or enrichment component can be started. We proceed with the cleaning component, which mainly deals with (a) the verification of the KG against a set of constraints and (b) the validation of each statement in the KG. The outcomes of this component are the verification and validation report.
Third, the enrichment component aims to detect duplicates in the KG. Therefore, it runs the instance matching process helped with a configuration learning module, which tries to find a tuned configuration. The resulting duplicates report triggers the entity fusion process supported by fusion strategies based on the assessment report of the KGs.
Finally, after having curated a KG, it is possible to repeat the curation process of the KG.

We start this section by introducing quality dimensions for KG assessment (Section~\ref{subsec:kg-assessment}). Later, we describe the KG cleaning (Section~\ref{subsec:kg-cleaning}) tasks, as well as KG enrichment (Section~\ref{subsec:kg-enrichment}) tasks.

\subsection{Knowledge Graph Assessment}
\label{subsec:kg-assessment}
Assessing the status of KGs is the first step to curate KGs. In recent years, several KGs have been created and released as open (e.g. DBpedia, Wikidata) or proprietary (e.g. Amazon Product Knowledge Graph).
We observe that these KGs widely vary in their quality, from manually curated KGs to automatically extracted KGs. Therefore, a data consumer needs to face the challenge to define a useful data source for specific tasks (i.e. ``fitness for use''). There are a number of studies, which have identified data quality dimension into various categories \cite{BatiniCFM09,BatiniS06,FarberBMR18,FenselSAHKPTUW20,PipinoLW02,Wang98,WangZL01,ZaveriRMPLA16} with the aim of measuring the usefulness of knowledge sources. For instance,~\cite{BatiniCFM09} describe a comparative analysis of methodologies and strategies of data quality dimensions, and~\cite{FarberBMR18} adopt some of these criteria to compare several data sources such as Freebase, OpenCyc, Wikidata, and YAGO.

Based on the analysis of the works mentioned above related to data quality assessment, we summarize 20 quality dimensions to consider on assessing the status of a Knowledge Graph (KG):

\begin{enumerate}
    \item \textbf{Accessibility} implies that the KG must be available, provide a public SPARQL endpoint, retrievable in RDF format, support content negotiation, and describes a licence.
    \item \textbf{Accuracy} defines the syntactic and semantic validity of assertions contained in the KG.
    \item \textbf{Appropriate amount} evaluates whether the KG contains knowledge for specific use case scenarios.
    \item \textbf{Believability} or trustworthiness measures whether the KG provides provenance information, and it is verifiable.
    \item \textbf{Completeness} in terms of schema and instance level for a specific use case.
    \item \textbf{Concise representation} evaluates the use of blank nodes and reification.
    \item \textbf{Consistent representation} detects the existence of disjoint inconsistencies of classes and schema restrictions in the KG.
    \item \textbf{Cost-effectiveness} measures the degree to which accurate data is necessary.
    \item \textbf{Ease of manipulation} evaluates the existence of documentation for manipulating the knowledge contained in the KG.
    \item \textbf{Ease of operation} refers to the possibility of updating, downloading, and integrating the KG. 
    \item \textbf{Ease of understanding} evaluates whether self-descriptive URIs are used and knowledge is presented in more than one language.
    \item \textbf{Free-of-error} refers to the total number of wrong and missing assertions contained in the KG.
    \item \textbf{Interoperability} evaluates whether the KG re-uses standard vocabularies and complies with Linked Open Data 5 Star.
    \item \textbf{Objectivity} defines the degree to which the KG is unbiased and impartial.
    \item \textbf{Relevancy} evaluates the level of applicability, in terms of domain coverage, of the KG to a specific use case.
    \item \textbf{Reputation} measures whether exist explicit trust ratings to the KG or there exist qualifiers on KG's statements.
    \item \textbf{Security} evaluates the degree to which the KG uses a digital signature and verifies the identity of the publisher.
    \item \textbf{Timeliness} measures the frequency of updates occurring in the KG and the validity period of its statements.
    \item \textbf{Traceability} evaluates the degree to which the KG provides provenance information and keeps a log of edits and changes.
    \item \textbf{Variety} refers to the degree to which the KG contains knowledge from different sources and various domains.
\end{enumerate}

The approach provides a practical framework for effectively assessing the status of KGs. Additionally, different quality dimensions may have different degrees of importance for different application scenarios. For instance, the Timeliness dimension may be very important in a domain that has predominantly dynamic data. Therefore, we let users define the weight of importance for each quality dimension. The tools proposed in~\cite{FenselSAHKPTUW20,Paulheim17,SimsekAKOSUF21,VaidyambathDSB19} can be used for measuring the proposed quality dimensions in this paper.

\subsection{Knowledge Graph Cleaning}
\label{subsec:kg-cleaning}
This task aims to improve the correctness of the KGs, which may contain a significant amount of syntax and semantic errors. For that, we distinguish between the verification and validation of KGs. The first aims to evaluate schema conformance and integrity constraints of KGs. The second one checks whether KGs accurately describe or represent the so-called ``real'' world.

\subsubsection{Verification.}
\label{subsubsec:kg-verification}
It is the process of evaluating KGs with formal specifications of integrity constraints. In a heterogeneous environment of structured data like KGs, there is not necessarily a unique constraint language for verifying KGs. We distinguish three categories:
\begin{enumerate}
    \item[i] \textbf{Query-based approaches}, such as Schemarama \cite{schemarama2001} that applies the XPath method and uses SquisQL language, and SPARQL Query Language\footnote{\url{https://www.w3.org/TR/sparql11-query/}}, Simple Application-Specific Constraints \cite{Simister2013}, SPARQL Inferencing Notation (SPIN), RDFUnit \cite{kontokostas2014test}, Shape Expressions (ShEx), and Shapes Constraint Language (SHACL) that use SPARQL queries like ASK and CONSTRUCT.
    \item[ii] \textbf{Inference-based approaches}, such as TreeHugger~\cite{TreeHugger2004} and Schematron rules that use XPath, and Stardog ICV \cite{Cerans2012} that implements subsumption, domain-range, participation, cardinality, and property constraints. 
    \item[iii] \textbf{Structural languages}, such as OSLC Resource Shape \cite{Ryman2013} that defines property constraints, Dublin Core Application Profiles \cite{Coyle2013} that defines constraints of values and cardinality of properties, and RDF Data Description (RDD) \cite{Fischer2015} that defines property and class constraints.
\end{enumerate}
We have presented constraint languages for verifying KGs. We noticed that some constraints are easier to write in one syntax (e.g. ShEx) than in others (e.g. SHACL). Furthermore, we distinguished between query-based, inference-based, and structural language approaches. These approaches can be used for the verification of KGs.

\subsubsection{Validation.}
\label{subsubsec:kg-validation}
It is a critical task to provide accurate, correct, and reliable knowledge. The knowledge validation task in KGs evaluates whether an assertion (e.g., ``Bill Gates is 64 years old'') from a KG is semantically correct or not and whether it corresponds with the so-called ``real'' world. We surveyed methods for validating KGs, we distinguish them according to the data used by them:
\begin{enumerate}
    \item[i]\textbf{Internal approaches} rely on statements or triples that exist within a KG. For instance, some methods identify statements as pieces of evidence to support a particular statement~\cite{JiaXCWE19,LiLXZ15,ShiW16,ShiralkarFMC2017,SyedRN2019}. Moreover, there exist approaches that use outlier detection techniques to evaluate whether a property value is out of the assumed distribution \cite{SyedRN2019,ThorneV17,WienandP2014} and approaches that use embedding models~\cite{LiLL20}. Furthermore, we can mention some tools, like COPAAL~\cite{SyedRN2019} and KGTtm~\cite{JiaXCWE19} that evaluate possible interesting relationship between entity pairs (subject,object) within a KG. 
    \item[ii] \textbf{External approaches} use external sources like the Freebase source to validate a statement. For example, there are approaches that use websites information~\cite{DongGHHLMSSZ14,GerberELBUNS2015,SpeckN19}, Linked Open Data datasets~\cite{SyedRN2018}, Wikipedia pages~\cite{ErcanEH19,PadiaFF18}, and DBpedia knowledge base~\cite{HuamanTBF2021,Rula2019}. Furthermore, there are methods that use topic coherence \cite{AletrasS13} and information extraction \cite{SpeckN19} techniques to validate KGs. The proposed tools are DeFacto~\cite{GerberELBUNS2015}, ExFaKT~\cite{GadElrab0UW2019}, Leopard~\cite{SpeckN19}, FactCheck~\cite{SyedRN2018}, and FacTify~\cite{ErcanEH19}, which rely on the Web and/or external knowledge sources like Wikipedia.
\end{enumerate}
The reviewed approaches are mostly focused on validating well-disseminated knowledge than factual knowledge, e.g. Wikipedia is the most frequently used by external approaches. The approaches mentioned above can validate KGs.

\subsection{Knowledge Graph Enrichment}
\label{subsec:kg-enrichment}
Enriching KGs is a process of high practical relevance to improve the completeness of KGs, and there is a need for effective\footnote{Achieving the comparison of all records.} and efficient\footnote{Optimizing the speed and used resources to compare a large number of records.} frameworks to tackle the problem. For doing that, we identified two tasks:
\begin{enumerate}
    \item[i] \textbf{Identifying and resolving duplicates} is identifying where two or more records in a single or various KGs are referring to the same entity and linking those. The tools found during the review of the literature are ADEL \cite{PluTR2017}, DDaaS~\cite{SimsekAKOSUF21}, Dedupe \cite{BilenkoM2003}, DuDe \cite{Draisbach2010}, Duke \cite{GarsholB2013}, Legato \cite{AchichiBT2017}, LIMES \cite{NgomoA2011}, SERIMI \cite{AraujoHSV2011}, and Silk \cite{VolzBGK2009}.
    \item[ii] \textbf{Resolving conflicting property value assertions} or data fusion refers to handle for example situations such as the pair of duplicated entities have a different value for the same property, the state-of-the-art tools for tackling this task are FAGI \cite{GiannopoulosSMKA2014}, Sieve \cite{MendesMB2012}, and SLIPO Toolkit \cite{AthanasiouAGKKM2019}.
\end{enumerate}
Most of the tools mentioned above need a previous configuration to start working, such as Silk and Sieve. Also, most of the approaches focus on an individual type of use case (e.g. FAGI focuses on geospatial data). We also notice that these tools are mostly focused on the detection of duplicates rather than on the resolution of the conflicting property values. It is important to note that when we resolve property value conflicts from different KGs, we need to assess them in order to know which KG is reliable and suitable for the task at hand. Besides, the identification of new relevant KGs must be done in advance.
\section{Discussion and Findings}
\label{sec:findings}
From the description of our framework in Section~\ref{sec:framework}, we can notice that is numerous approaches proposed for improving the quality of KGs, either for assessing the status of KGs, for detecting and correcting errors, or for detecting duplicates and perform knowledge fusion. We discuss our findings as follows:
\begin{itemize}
    \item \textbf{Automation}. Various quality dimensions can hardly be fully automated for a technical or operational reason. Furthermore, It is desirable to allow users to create a semi-automatic \textbf{mapping} (or schema alignment~\cite{DongS2015}) between their KG and another KG.
    \item \textbf{Cost-effectiveness}. Validating KGs may lead to a high cost of deployment, due to its dependency on proprietary services (e.g., search engines). This can be overcome, to a certain level, if a validation framework uses open corpora (e.g. Wikipedia) but its performance lows down.
    \item \textbf{Dynamic data}. We should add the complexity of dynamic data (i.e., fast-changing data) since statements can be represented differently over a period of time. For instance, the telephone number of a restaurant can change.
    \item \textbf{Prevention}. Fixing syntactic and semantic errors can be caught during the creation and hosting of KGs. For instance, checking whether the input data conform to a specific schema. Furthermore, we have observed that the expressivity of constraint languages is directly related with the expressivity of SPARQL.
    \item \textbf{Reproducibility}. On one hand, most of the tools provide bast documentation, on the other hand, we notice usability issues of tools, e.g., complex to apply in different domains. Also, many of them were abandoned in their GitHub repositories and no longer maintained.
    \item \textbf{Re-usability}. Knowledge assessment frameworks may help to identify reliable and trustworthy KGs, to which user's KG can be interlinked, e.g., the quality assessment may help to define to which extent a KG can be used for interlinking entities. Furthermore, most of the duplicate detection frameworks offer only simple similarity metrics, however in complex cases, complex metrics are needed.
    \item \textbf{User-in-the-Loop}. Users can define the degree of importance of a quality dimension (i.e. weights), for assessing a KG, according to the task at hand. For instance, knowledge assessment can help to decide which KG is best for resolving conflicting property values.
    \item \textbf{Scalability}. Existing frameworks are still lacking scalability to large KGs. For instance, applying a genetic algorithm can automatically tune a configuration for duplicate detection (i.e. Configuration learning process).
    \item \textbf{Trade-off between completeness and correctness}. Most of the approaches only detect errors or missing values but leave the correction part completely to users.
\end{itemize}
Our framework described in Section~\ref{sec:framework} aims to provide a practical curation framework that can be used for improving the quality of KGs. Above, we listed our findings that one can consider on the development of future curation frameworks.
\section{Conclusion and Future Work}
\label{sec:conclusion}
Although our paper has presented methods and tools, improvement suggestions, and workflow in knowledge graph curation, we believe that there is still work to do in this field. This work aimed to fill this gap and facilitate future research in KGs curation domain. Twenty dimensions of KG quality were explored, several tools were listed for the cleaning and enrichment of KGs. Furthermore, this paper presents building modules that KGs architects can take into account in the development of future knowledge graph curation frameworks.

In the following, we point out our future work and open research questions. Firstly, our next steps involve the development of the KG curation framework to tackle the assessment, cleaning, and enrichment of KGs. Moreover, we will evaluate the performance of the framework and conduct surveys from domain experts and KG researchers to evaluate and improve the proposed knowledge curation framework.

\begin{acks}
I would like to thank all participants of the industrial research project MindLab (https://mindlab.ai/).
\end{acks}

\bibliographystyle{ACM-Reference-Format}
\bibliography{sample-base}


\begin{thebibliography}{51}


\ifx \showCODEN    \undefined \def \showCODEN     #1{\unskip}     \fi
\ifx \showDOI      \undefined \def \showDOI       #1{#1}\fi
\ifx \showISBNx    \undefined \def \showISBNx     #1{\unskip}     \fi
\ifx \showISBNxiii \undefined \def \showISBNxiii  #1{\unskip}     \fi
\ifx \showISSN     \undefined \def \showISSN      #1{\unskip}     \fi
\ifx \showLCCN     \undefined \def \showLCCN      #1{\unskip}     \fi
\ifx \shownote     \undefined \def \shownote      #1{#1}          \fi
\ifx \showarticletitle \undefined \def \showarticletitle #1{#1}   \fi
\ifx \showURL      \undefined \def \showURL       {\relax}        \fi
\providecommand\bibfield[2]{#2}
\providecommand\bibinfo[2]{#2}
\providecommand\natexlab[1]{#1}
\providecommand\showeprint[2][]{arXiv:#2}

\bibitem[\protect\citeauthoryear{Achichi, Bellahsene, and Todorov}{Achichi
  et~al\mbox{.}}{2017}]%
        {AchichiBT2017}
\bibfield{author}{\bibinfo{person}{Manel Achichi}, \bibinfo{person}{Zohra
  Bellahsene}, {and} \bibinfo{person}{Konstantin Todorov}.}
  \bibinfo{year}{2017}\natexlab{}.
\newblock \showarticletitle{Legato results for {OAEI} 2017}. In
  \bibinfo{booktitle}{\emph{Proceedings of the 16th International Semantic Web
  Conference {(ISWC2017)}: 12th Workshop on Ontology Matching {(OM2017)},
  Vienna, Austria, October 21, 2017}} \emph{(\bibinfo{series}{{CEUR} Workshop
  Proceedings}, Vol.~\bibinfo{volume}{2032})}.
  \bibinfo{publisher}{CEUR-WS.org}, \bibinfo{pages}{146--152}.
\newblock


\bibitem[\protect\citeauthoryear{Aletras and Stevenson}{Aletras and
  Stevenson}{2013}]%
        {AletrasS13}
\bibfield{author}{\bibinfo{person}{Nikolaos Aletras} {and}
  \bibinfo{person}{Mark Stevenson}.} \bibinfo{year}{2013}\natexlab{}.
\newblock \showarticletitle{Evaluating Topic Coherence Using Distributional
  Semantics}. In \bibinfo{booktitle}{\emph{Proceedings of the 10th
  International Conference on Computational Semantics, {(IWCS2013)}, Potsdam,
  Germany, March 19-22, 2013}}. \bibinfo{publisher}{The Association for
  Computer Linguistics}, \bibinfo{pages}{13--22}.
\newblock


\bibitem[\protect\citeauthoryear{Ara{\'{u}}jo, Hidders, Schwabe, and
  de~Vries}{Ara{\'{u}}jo et~al\mbox{.}}{2011}]%
        {AraujoHSV2011}
\bibfield{author}{\bibinfo{person}{Samur Ara{\'{u}}jo}, \bibinfo{person}{Jan
  Hidders}, \bibinfo{person}{Daniel Schwabe}, {and} \bibinfo{person}{Arjen~P.
  de Vries}.} \bibinfo{year}{2011}\natexlab{}.
\newblock \showarticletitle{{SERIMI} - Resource Description Similarity, {RDF}
  Instance Matching and Interlinking}. In \bibinfo{booktitle}{\emph{Proceedings
  of the 6th International Workshop on Ontology Matching {(OM2011)}, Bonn,
  Germany, October 24, 2011}} \emph{(\bibinfo{series}{{CEUR} Workshop
  Proceedings}, Vol.~\bibinfo{volume}{814})}. \bibinfo{publisher}{CEUR-WS.org}.
\newblock


\bibitem[\protect\citeauthoryear{Athanasiou, Alexakis, Giannopoulos,
  Karagiannakis, Kouvaras, Mitropoulos, Patroumpas, and Skoutas}{Athanasiou
  et~al\mbox{.}}{2019}]%
        {AthanasiouAGKKM2019}
\bibfield{author}{\bibinfo{person}{Spiros Athanasiou}, \bibinfo{person}{Michail
  Alexakis}, \bibinfo{person}{Giorgos Giannopoulos}, \bibinfo{person}{Nikos
  Karagiannakis}, \bibinfo{person}{Yannis Kouvaras}, \bibinfo{person}{Pantelis
  Mitropoulos}, \bibinfo{person}{Kostas Patroumpas}, {and}
  \bibinfo{person}{Dimitrios Skoutas}.} \bibinfo{year}{2019}\natexlab{}.
\newblock \showarticletitle{{SLIPO:} Large-Scale Data Integration for Points of
  Interest}. In \bibinfo{booktitle}{\emph{Proceedings of the 22nd International
  Conference on Extending Database Technology {(EDBT2019)}, Lisbon, Portugal,
  March 26-29, 2019}}. \bibinfo{publisher}{OpenProceedings.org},
  \bibinfo{pages}{574--577}.
\newblock


\bibitem[\protect\citeauthoryear{Batini, Cappiello, Francalanci, and
  Maurino}{Batini et~al\mbox{.}}{2009}]%
        {BatiniCFM09}
\bibfield{author}{\bibinfo{person}{Carlo Batini}, \bibinfo{person}{Cinzia
  Cappiello}, \bibinfo{person}{Chiara Francalanci}, {and}
  \bibinfo{person}{Andrea Maurino}.} \bibinfo{year}{2009}\natexlab{}.
\newblock \showarticletitle{Methodologies for data quality assessment and
  improvement}.
\newblock \bibinfo{journal}{\emph{{ACM} Comput. Surv.}} \bibinfo{volume}{41},
  \bibinfo{number}{3} (\bibinfo{year}{2009}), \bibinfo{pages}{16:1--16:52}.
\newblock


\bibitem[\protect\citeauthoryear{Batini and Scannapieco}{Batini and
  Scannapieco}{2006}]%
        {BatiniS06}
\bibfield{author}{\bibinfo{person}{Carlo Batini} {and} \bibinfo{person}{Monica
  Scannapieco}.} \bibinfo{year}{2006}\natexlab{}.
\newblock \bibinfo{booktitle}{\emph{Data Quality: Concepts, Methodologies and
  Techniques}}.
\newblock \bibinfo{publisher}{Springer}.
\newblock
\showISBNx{978-3-540-33172-8}


\bibitem[\protect\citeauthoryear{Bilenko and Mooney}{Bilenko and
  Mooney}{2003}]%
        {BilenkoM2003}
\bibfield{author}{\bibinfo{person}{Mikhail Bilenko} {and}
  \bibinfo{person}{Raymond~J. Mooney}.} \bibinfo{year}{2003}\natexlab{}.
\newblock \showarticletitle{Adaptive duplicate detection using learnable string
  similarity measures}. In \bibinfo{booktitle}{\emph{Proceedings of the 9th
  {ACM} International Conference on Knowledge Discovery and Data Mining
  {(SIGKDD2003)}, Washington, USA, August 24 - 27, 2003}}.
  \bibinfo{publisher}{{ACM}}, \bibinfo{pages}{39--48}.
\newblock


\bibitem[\protect\citeauthoryear{Cerans, Barzdins, Liepins, Ovcinnikova,
  Rikacovs, and Sprogis}{Cerans et~al\mbox{.}}{2012}]%
        {Cerans2012}
\bibfield{author}{\bibinfo{person}{Karlis Cerans}, \bibinfo{person}{Guntis
  Barzdins}, \bibinfo{person}{Renars Liepins}, \bibinfo{person}{Julija
  Ovcinnikova}, \bibinfo{person}{Sergejs Rikacovs}, {and}
  \bibinfo{person}{Arturs Sprogis}.} \bibinfo{year}{2012}\natexlab{}.
\newblock \showarticletitle{Graphical Schema Editing for Stardog OWL/RDF
  Databases using OWLGrEd/S.}. In \bibinfo{booktitle}{\emph{OWLED}},
  Vol.~\bibinfo{volume}{849}.
\newblock


\bibitem[\protect\citeauthoryear{Coyle and Baker}{Coyle and Baker}{2013}]%
        {Coyle2013}
\bibfield{author}{\bibinfo{person}{Karen Coyle} {and} \bibinfo{person}{Tom
  Baker}.} \bibinfo{year}{2013}\natexlab{}.
\newblock \showarticletitle{Dublin core application profiles. separating
  validation from semantics}. In \bibinfo{booktitle}{\emph{RDF Validation
  Workshop. Practical Assurances for Quality RDF Data, Cambridge, Ma, Boston}}.
\newblock
\urldef\tempurl%
\url{http://www.w3.org/2012/12/rdf-val}
\showURL{%
\tempurl}


\bibitem[\protect\citeauthoryear{Dong, Gabrilovich, Heitz, Horn, Lao, Murphy,
  Strohmann, Sun, and Zhang}{Dong et~al\mbox{.}}{2014}]%
        {DongGHHLMSSZ14}
\bibfield{author}{\bibinfo{person}{Xin Dong}, \bibinfo{person}{Evgeniy
  Gabrilovich}, \bibinfo{person}{Geremy Heitz}, \bibinfo{person}{Wilko Horn},
  \bibinfo{person}{Ni Lao}, \bibinfo{person}{Kevin Murphy},
  \bibinfo{person}{Thomas Strohmann}, \bibinfo{person}{Shaohua Sun}, {and}
  \bibinfo{person}{Wei Zhang}.} \bibinfo{year}{2014}\natexlab{}.
\newblock \showarticletitle{Knowledge vault: a web-scale approach to
  probabilistic knowledge fusion}. In \bibinfo{booktitle}{\emph{The 20th {ACM}
  {SIGKDD} International Conference on Knowledge Discovery and Data Mining,
  {KDD} '14, New York, NY, {USA} - August 24 - 27, 2014}}.
  \bibinfo{publisher}{{ACM}}, \bibinfo{pages}{601--610}.
\newblock


\bibitem[\protect\citeauthoryear{Dong}{Dong}{2019}]%
        {Dong2019}
\bibfield{author}{\bibinfo{person}{Xin~Luna Dong}.}
  \bibinfo{year}{2019}\natexlab{}.
\newblock \showarticletitle{Building a Broad Knowledge Graph for Products}. In
  \bibinfo{booktitle}{\emph{35th {IEEE} International Conference on Data
  Engineering, {ICDE} 2019, Macao, China, April 8-11, 2019}}.
  \bibinfo{publisher}{{IEEE}}, \bibinfo{pages}{25}.
\newblock
\urldef\tempurl%
\url{https://doi.org/10.1109/ICDE.2019.00010}
\showDOI{\tempurl}


\bibitem[\protect\citeauthoryear{Dong and Srivastava}{Dong and
  Srivastava}{2015}]%
        {DongS2015}
\bibfield{author}{\bibinfo{person}{Xin~Luna Dong} {and} \bibinfo{person}{Divesh
  Srivastava}.} \bibinfo{year}{2015}\natexlab{}.
\newblock \bibinfo{booktitle}{\emph{Big Data Integration}}.
\newblock \bibinfo{publisher}{Morgan {\&} Claypool Publishers}.
\newblock


\bibitem[\protect\citeauthoryear{Draisbach and Naumann}{Draisbach and
  Naumann}{2008}]%
        {Draisbach2010}
\bibfield{author}{\bibinfo{person}{Uwe Draisbach} {and} \bibinfo{person}{Felix
  Naumann}.} \bibinfo{year}{2008}\natexlab{}.
\newblock \showarticletitle{DuDe: The duplicate detection toolkit}. In
  \bibinfo{booktitle}{\emph{Proceedings of the 36th Conference on Very Large
  Data Bases {(VLDB2010)}: Workshop on Quality in Databases, Singapore,
  Singapore, September 13 - 17, 2010}}.
\newblock


\bibitem[\protect\citeauthoryear{Ercan, Elbassuoni, and Hose}{Ercan
  et~al\mbox{.}}{2019}]%
        {ErcanEH19}
\bibfield{author}{\bibinfo{person}{Gonenc Ercan}, \bibinfo{person}{Shady
  Elbassuoni}, {and} \bibinfo{person}{Katja Hose}.}
  \bibinfo{year}{2019}\natexlab{}.
\newblock \showarticletitle{Retrieving Textual Evidence for Knowledge Graph
  Facts}. In \bibinfo{booktitle}{\emph{Proceedings of the 16th European
  Semantic Web Conference {(ESWC 2019)}, Portoro{\v{z}}, Slovenia, June 2-6,
  2019}} \emph{(\bibinfo{series}{Lecture Notes in Computer Science},
  Vol.~\bibinfo{volume}{11503})}. \bibinfo{publisher}{Springer},
  \bibinfo{pages}{52--67}.
\newblock


\bibitem[\protect\citeauthoryear{F{\"{a}}rber, Bartscherer, Menne, and
  Rettinger}{F{\"{a}}rber et~al\mbox{.}}{2018}]%
        {FarberBMR18}
\bibfield{author}{\bibinfo{person}{Michael F{\"{a}}rber},
  \bibinfo{person}{Frederic Bartscherer}, \bibinfo{person}{Carsten Menne},
  {and} \bibinfo{person}{Achim Rettinger}.} \bibinfo{year}{2018}\natexlab{}.
\newblock \showarticletitle{Linked data quality of DBpedia, Freebase, OpenCyc,
  Wikidata, and {YAGO}}.
\newblock \bibinfo{journal}{\emph{Semantic Web}} \bibinfo{volume}{9},
  \bibinfo{number}{1} (\bibinfo{year}{2018}), \bibinfo{pages}{77--129}.
\newblock


\bibitem[\protect\citeauthoryear{Fensel, Simsek, Angele, Huaman, K{\"{a}}rle,
  Panasiuk, Toma, Umbrich, and Wahler}{Fensel et~al\mbox{.}}{2020}]%
        {FenselSAHKPTUW20}
\bibfield{author}{\bibinfo{person}{Dieter Fensel}, \bibinfo{person}{Umutcan
  Simsek}, \bibinfo{person}{Kevin Angele}, \bibinfo{person}{Elwin Huaman},
  \bibinfo{person}{Elias K{\"{a}}rle}, \bibinfo{person}{Oleksandra Panasiuk},
  \bibinfo{person}{Ioan Toma}, \bibinfo{person}{J{\"{u}}rgen Umbrich}, {and}
  \bibinfo{person}{Alexander Wahler}.} \bibinfo{year}{2020}\natexlab{}.
\newblock \bibinfo{booktitle}{\emph{Knowledge Graphs - Methodology, Tools and
  Selected Use Cases}}.
\newblock \bibinfo{publisher}{Springer}.
\newblock
\showISBNx{978-3-030-37438-9}
\urldef\tempurl%
\url{https://doi.org/10.1007/978-3-030-37439-6}
\showDOI{\tempurl}


\bibitem[\protect\citeauthoryear{Fischer, Lausen, Sch{\"{a}}tzle, and
  Schmidt}{Fischer et~al\mbox{.}}{2015}]%
        {Fischer2015}
\bibfield{author}{\bibinfo{person}{Peter~M. Fischer}, \bibinfo{person}{Georg
  Lausen}, \bibinfo{person}{Alexander Sch{\"{a}}tzle}, {and}
  \bibinfo{person}{Michael Schmidt}.} \bibinfo{year}{2015}\natexlab{}.
\newblock \showarticletitle{{RDF} Constraint Checking}. In
  \bibinfo{booktitle}{\emph{Proceedings of the Workshops of the 2015 Joint
  Conference {(EDBT/ICDT)}}} \emph{(\bibinfo{series}{{CEUR} Workshop
  Proceedings}, Vol.~\bibinfo{volume}{1330})}.
  \bibinfo{publisher}{CEUR-WS.org}, \bibinfo{pages}{205--212}.
\newblock


\bibitem[\protect\citeauthoryear{Gad{-}Elrab, Stepanova, Urbani, and
  Weikum}{Gad{-}Elrab et~al\mbox{.}}{2019}]%
        {GadElrab0UW2019}
\bibfield{author}{\bibinfo{person}{Mohamed~H. Gad{-}Elrab},
  \bibinfo{person}{Daria Stepanova}, \bibinfo{person}{Jacopo Urbani}, {and}
  \bibinfo{person}{Gerhard Weikum}.} \bibinfo{year}{2019}\natexlab{}.
\newblock \showarticletitle{ExFaKT: {A} Framework for Explaining Facts over
  Knowledge Graphs and Text}. In \bibinfo{booktitle}{\emph{Proceedings of the
  12th {ACM} International Conference on Web Search and Data Mining,
  {(WSDM2019)}, Melbourne, Australia, February 11-15, 2019}}.
  \bibinfo{publisher}{{ACM}}, \bibinfo{pages}{87--95}.
\newblock


\bibitem[\protect\citeauthoryear{Garshol and Borge}{Garshol and Borge}{2013}]%
        {GarsholB2013}
\bibfield{author}{\bibinfo{person}{Lars~Marius Garshol} {and}
  \bibinfo{person}{Axel Borge}.} \bibinfo{year}{2013}\natexlab{}.
\newblock \showarticletitle{Hafslund Sesam - An Archive on Semantics}. In
  \bibinfo{booktitle}{\emph{Proceedings of the 10th Extending Semantic Web
  Conference (ESWC2013), Montpellier, France, May 26-30, 2013}}
  \emph{(\bibinfo{series}{LNCS}, Vol.~\bibinfo{volume}{7882})}.
  \bibinfo{publisher}{Springer}, \bibinfo{pages}{578--592}.
\newblock


\bibitem[\protect\citeauthoryear{Gerber, Esteves, Lehmann, B{\"{u}}hmann,
  Usbeck, Ngomo, and Speck}{Gerber et~al\mbox{.}}{2015}]%
        {GerberELBUNS2015}
\bibfield{author}{\bibinfo{person}{Daniel Gerber}, \bibinfo{person}{Diego
  Esteves}, \bibinfo{person}{Jens Lehmann}, \bibinfo{person}{Lorenz
  B{\"{u}}hmann}, \bibinfo{person}{Ricardo Usbeck},
  \bibinfo{person}{Axel{-}Cyrille~Ngonga Ngomo}, {and}
  \bibinfo{person}{Ren{\'{e}} Speck}.} \bibinfo{year}{2015}\natexlab{}.
\newblock \showarticletitle{DeFacto - Temporal and multilingual Deep Fact
  Validation}.
\newblock \bibinfo{journal}{\emph{Journal of Web Semantics}}
  \bibinfo{volume}{35} (\bibinfo{year}{2015}), \bibinfo{pages}{85--101}.
\newblock


\bibitem[\protect\citeauthoryear{Giannopoulos, Skoutas, Maroulis,
  Karagiannakis, and Athanasiou}{Giannopoulos et~al\mbox{.}}{2014}]%
        {GiannopoulosSMKA2014}
\bibfield{author}{\bibinfo{person}{Giorgos Giannopoulos},
  \bibinfo{person}{Dimitrios Skoutas}, \bibinfo{person}{Thomas Maroulis},
  \bibinfo{person}{Nikos Karagiannakis}, {and} \bibinfo{person}{Spiros
  Athanasiou}.} \bibinfo{year}{2014}\natexlab{}.
\newblock \showarticletitle{{FAGI:} {A} Framework for Fusing Geospatial {RDF}
  Data}. In \bibinfo{booktitle}{\emph{Proceedings of the Confederated
  International Conferences "On the Move to Meaningful Internet Systems"
  (OTM2014), Amantea, Italy, October 27-31, 2014}}
  \emph{(\bibinfo{series}{LNCS}, Vol.~\bibinfo{volume}{8841})}.
  \bibinfo{publisher}{Springer}, \bibinfo{pages}{553--561}.
\newblock


\bibitem[\protect\citeauthoryear{Huaman, Tauqeer, Bushati, and Fensel}{Huaman
  et~al\mbox{.}}{2021}]%
        {HuamanTBF2021}
\bibfield{author}{\bibinfo{person}{Elwin Huaman}, \bibinfo{person}{Amar
  Tauqeer}, \bibinfo{person}{Geni Bushati}, {and} \bibinfo{person}{Anna
  Fensel}.} \bibinfo{year}{2021}\natexlab{}.
\newblock \showarticletitle{Towards Knowledge Graphs Validation through
  Weighted Knowledge Sources}.
\newblock \bibinfo{journal}{\emph{CoRR}}  \bibinfo{volume}{abs/2104.12622}
  (\bibinfo{year}{2021}).
\newblock
\showeprint[arXiv]{2104.12622}
\urldef\tempurl%
\url{https://arxiv.org/abs/2104.12622}
\showURL{%
\tempurl}


\bibitem[\protect\citeauthoryear{Jia, Xiang, Chen, Wang, and E}{Jia
  et~al\mbox{.}}{2019}]%
        {JiaXCWE19}
\bibfield{author}{\bibinfo{person}{Shengbin Jia}, \bibinfo{person}{Yang Xiang},
  \bibinfo{person}{Xiaojun Chen}, \bibinfo{person}{Kun Wang}, {and}
  \bibinfo{person}{Shijia E}.} \bibinfo{year}{2019}\natexlab{}.
\newblock \showarticletitle{Triple Trustworthiness Measurement for Knowledge
  Graph}. In \bibinfo{booktitle}{\emph{Proceedings of The World Wide Web
  Conference, {(WWW2019)}, San Francisco, USA, May 13-17, 2019}}.
  \bibinfo{publisher}{{ACM}}, \bibinfo{pages}{2865--2871}.
\newblock


\bibitem[\protect\citeauthoryear{Kontokostas, Westphal, Auer, Hellmann,
  Lehmann, Cornelissen, and Zaveri}{Kontokostas et~al\mbox{.}}{2014}]%
        {kontokostas2014test}
\bibfield{author}{\bibinfo{person}{Dimitris Kontokostas},
  \bibinfo{person}{Patrick Westphal}, \bibinfo{person}{S{\"o}ren Auer},
  \bibinfo{person}{Sebastian Hellmann}, \bibinfo{person}{Jens Lehmann},
  \bibinfo{person}{Roland Cornelissen}, {and} \bibinfo{person}{Amrapali
  Zaveri}.} \bibinfo{year}{2014}\natexlab{}.
\newblock \showarticletitle{Test-driven evaluation of linked data quality}. In
  \bibinfo{booktitle}{\emph{Proceedings of the 23rd international conference on
  World Wide Web}}. ACM, \bibinfo{pages}{747--758}.
\newblock


\bibitem[\protect\citeauthoryear{Li, Li, Xu, and Zhong}{Li
  et~al\mbox{.}}{2015}]%
        {LiLXZ15}
\bibfield{author}{\bibinfo{person}{Huiying Li}, \bibinfo{person}{Yuanyuan Li},
  \bibinfo{person}{Feifei Xu}, {and} \bibinfo{person}{Xinyu Zhong}.}
  \bibinfo{year}{2015}\natexlab{}.
\newblock \showarticletitle{Probabilistic Error Detecting in Numerical Linked
  Data}. In \bibinfo{booktitle}{\emph{Database and Expert Systems Applications
  - 26th International Conference, {DEXA} 2015, Valencia, Spain, September 1-4,
  2015, Proceedings, Part {I}}} \emph{(\bibinfo{series}{Lecture Notes in
  Computer Science}, Vol.~\bibinfo{volume}{9261})}.
  \bibinfo{publisher}{Springer}, \bibinfo{pages}{61--75}.
\newblock


\bibitem[\protect\citeauthoryear{Li, Li, and Lei}{Li et~al\mbox{.}}{2020}]%
        {LiLL20}
\bibfield{author}{\bibinfo{person}{Yunfeng Li}, \bibinfo{person}{Xiaoyong Li},
  {and} \bibinfo{person}{Mingjian Lei}.} \bibinfo{year}{2020}\natexlab{}.
\newblock \showarticletitle{CTransE: An Effective Information Credibility
  Evaluation Method Based on Classified Translating Embedding in Knowledge
  Graphs}. In \bibinfo{booktitle}{\emph{Database and Expert Systems
  Applications - 31st International Conference, {DEXA} 2020, Bratislava,
  Slovakia, September 14-17, 2020, Proceedings, Part {II}}}
  \emph{(\bibinfo{series}{Lecture Notes in Computer Science},
  Vol.~\bibinfo{volume}{12392})}. \bibinfo{publisher}{Springer},
  \bibinfo{pages}{287--300}.
\newblock


\bibitem[\protect\citeauthoryear{Mendes, M{\"{u}}hleisen, and Bizer}{Mendes
  et~al\mbox{.}}{2012}]%
        {MendesMB2012}
\bibfield{author}{\bibinfo{person}{Pablo~N. Mendes}, \bibinfo{person}{Hannes
  M{\"{u}}hleisen}, {and} \bibinfo{person}{Christian Bizer}.}
  \bibinfo{year}{2012}\natexlab{}.
\newblock \showarticletitle{{Sieve: linked data quality assessment and
  fusion}}. In \bibinfo{booktitle}{\emph{Proceedings of 2nd International
  Workshop on Linked Web Data Management {(LWDM 2012)}, in conjunction with the
  15th International Conference on Extending Database Technology {(EDBT2012)}:
  Workshops, Berlin, Germany, March 30, 2012}}. \bibinfo{publisher}{{ACM}},
  \bibinfo{pages}{116--123}.
\newblock


\bibitem[\protect\citeauthoryear{Miller and Brickley}{Miller and
  Brickley}{2001}]%
        {schemarama2001}
\bibfield{author}{\bibinfo{person}{Libby Miller} {and} \bibinfo{person}{Dan
  Brickley}.} \bibinfo{year}{2001}\natexlab{}.
\newblock \showarticletitle{RDF: Schemarama}.
\newblock \bibinfo{journal}{\emph{ILRT}} (\bibinfo{year}{2001}).
\newblock
\urldef\tempurl%
\url{https://web.archive.org/web/20011119222635/}
\showURL{%
\tempurl}


\bibitem[\protect\citeauthoryear{Ngomo and Auer}{Ngomo and Auer}{2011}]%
        {NgomoA2011}
\bibfield{author}{\bibinfo{person}{Axel{-}Cyrille~Ngonga Ngomo} {and}
  \bibinfo{person}{S{\"{o}}ren Auer}.} \bibinfo{year}{2011}\natexlab{}.
\newblock \showarticletitle{{LIMES} - {A} Time-Efficient Approach for
  Large-Scale Link Discovery on the Web of Data}. In
  \bibinfo{booktitle}{\emph{Proceedings of the 22nd International Joint
  Conference on Artificial Intelligence {(IJCAI2011)}, Barcelona, Spain, July
  16–22, 2011}}. \bibinfo{publisher}{AAAI Press},
  \bibinfo{pages}{2312--2317}.
\newblock


\bibitem[\protect\citeauthoryear{Noy, Gao, Jain, Narayanan, Patterson, and
  Taylor}{Noy et~al\mbox{.}}{2019}]%
        {NoyGJNPT2019}
\bibfield{author}{\bibinfo{person}{Natasha~F. Noy}, \bibinfo{person}{Yuqing
  Gao}, \bibinfo{person}{Anshu Jain}, \bibinfo{person}{Anant Narayanan},
  \bibinfo{person}{Alan Patterson}, {and} \bibinfo{person}{Jamie Taylor}.}
  \bibinfo{year}{2019}\natexlab{}.
\newblock \showarticletitle{Industry-scale Knowledge Graphs: Lessons and
  Challenges}.
\newblock \bibinfo{journal}{\emph{{ACM} Queue}} \bibinfo{volume}{17},
  \bibinfo{number}{2} (\bibinfo{year}{2019}), \bibinfo{pages}{20}.
\newblock


\bibitem[\protect\citeauthoryear{Padia, Ferraro, and Finin}{Padia
  et~al\mbox{.}}{2018}]%
        {PadiaFF18}
\bibfield{author}{\bibinfo{person}{Ankur Padia}, \bibinfo{person}{Francis
  Ferraro}, {and} \bibinfo{person}{Tim Finin}.}
  \bibinfo{year}{2018}\natexlab{}.
\newblock \showarticletitle{{SURFACE:} Semantically Rich Fact Validation with
  Explanations}.
\newblock \bibinfo{journal}{\emph{CoRR}}  \bibinfo{volume}{abs/1810.13223}
  (\bibinfo{year}{2018}).
\newblock
\showeprint[arxiv]{1810.13223}


\bibitem[\protect\citeauthoryear{Paulheim}{Paulheim}{2017}]%
        {Paulheim17}
\bibfield{author}{\bibinfo{person}{Heiko Paulheim}.}
  \bibinfo{year}{2017}\natexlab{}.
\newblock \showarticletitle{Knowledge graph refinement: {A} survey of
  approaches and evaluation methods}.
\newblock \bibinfo{journal}{\emph{Semantic Web}} \bibinfo{volume}{8},
  \bibinfo{number}{3} (\bibinfo{year}{2017}), \bibinfo{pages}{489--508}.
\newblock


\bibitem[\protect\citeauthoryear{Pipino, Lee, and Wang}{Pipino
  et~al\mbox{.}}{2002}]%
        {PipinoLW02}
\bibfield{author}{\bibinfo{person}{Leo Pipino}, \bibinfo{person}{Yang~W. Lee},
  {and} \bibinfo{person}{Richard~Y. Wang}.} \bibinfo{year}{2002}\natexlab{}.
\newblock \showarticletitle{Data quality assessment}.
\newblock \bibinfo{journal}{\emph{Commun. {ACM}}} \bibinfo{volume}{45},
  \bibinfo{number}{4} (\bibinfo{year}{2002}), \bibinfo{pages}{211--218}.
\newblock


\bibitem[\protect\citeauthoryear{Plu, Troncy, and Rizzo}{Plu
  et~al\mbox{.}}{2017}]%
        {PluTR2017}
\bibfield{author}{\bibinfo{person}{Julien Plu}, \bibinfo{person}{Rapha{\"{e}}l
  Troncy}, {and} \bibinfo{person}{Giuseppe Rizzo}.}
  \bibinfo{year}{2017}\natexlab{}.
\newblock \showarticletitle{ADEL@OKE 2017: {A} Generic Method for Indexing
  Knowledge Bases for Entity Linking}. In \bibinfo{booktitle}{\emph{Proceedings
  of the 4th Semantic Web Evaluation Challenge at {ESWC2017}, Portoroz,
  Slovenia, May 28 - June 1, 2017}} \emph{(\bibinfo{series}{CCIS},
  Vol.~\bibinfo{volume}{769})}. \bibinfo{publisher}{Springer},
  \bibinfo{pages}{49--55}.
\newblock


\bibitem[\protect\citeauthoryear{Rula, Palmonari, Rubinacci, Ngomo, Lehmann,
  Maurino, and Esteves}{Rula et~al\mbox{.}}{2019}]%
        {Rula2019}
\bibfield{author}{\bibinfo{person}{Anisa Rula}, \bibinfo{person}{Matteo
  Palmonari}, \bibinfo{person}{Simone Rubinacci},
  \bibinfo{person}{Axel{-}Cyrille~Ngonga Ngomo}, \bibinfo{person}{Jens
  Lehmann}, \bibinfo{person}{Andrea Maurino}, {and} \bibinfo{person}{Diego
  Esteves}.} \bibinfo{year}{2019}\natexlab{}.
\newblock \showarticletitle{{{TISCO:} Temporal scoping of facts}}.
\newblock \bibinfo{journal}{\emph{Journal of Web Semantics}}
  \bibinfo{volume}{54} (\bibinfo{year}{2019}), \bibinfo{pages}{72--86}.
\newblock


\bibitem[\protect\citeauthoryear{Ryman, Le~Hors, and Speicher}{Ryman
  et~al\mbox{.}}{2013}]%
        {Ryman2013}
\bibfield{author}{\bibinfo{person}{Arthur~G Ryman}, \bibinfo{person}{Arnaud
  Le~Hors}, {and} \bibinfo{person}{Steve Speicher}.}
  \bibinfo{year}{2013}\natexlab{}.
\newblock \showarticletitle{OSLC Resource Shape: A language for defining
  constraints on Linked Data.}
\newblock \bibinfo{journal}{\emph{LDOW}}  \bibinfo{volume}{996}
  (\bibinfo{year}{2013}).
\newblock


\bibitem[\protect\citeauthoryear{Shi and Weninger}{Shi and Weninger}{2016}]%
        {ShiW16}
\bibfield{author}{\bibinfo{person}{Baoxu Shi} {and} \bibinfo{person}{Tim
  Weninger}.} \bibinfo{year}{2016}\natexlab{}.
\newblock \showarticletitle{Discriminative predicate path mining for fact
  checking in knowledge graphs}.
\newblock \bibinfo{journal}{\emph{Knowledge-Based Systems}}
  \bibinfo{volume}{104} (\bibinfo{year}{2016}), \bibinfo{pages}{123--133}.
\newblock


\bibitem[\protect\citeauthoryear{Shiralkar, Flammini, Menczer, and
  Ciampaglia}{Shiralkar et~al\mbox{.}}{2017}]%
        {ShiralkarFMC2017}
\bibfield{author}{\bibinfo{person}{Prashant Shiralkar},
  \bibinfo{person}{Alessandro Flammini}, \bibinfo{person}{Filippo Menczer},
  {and} \bibinfo{person}{Giovanni~Luca Ciampaglia}.}
  \bibinfo{year}{2017}\natexlab{}.
\newblock \showarticletitle{Finding Streams in Knowledge Graphs to Support Fact
  Checking}. In \bibinfo{booktitle}{\emph{Proceedings of the {IEEE}
  International Conference on Data Mining {(ICDM2017)}, New Orleans, USA,
  November 18-21, 2017}}. \bibinfo{publisher}{{IEEE} Computer Society},
  \bibinfo{pages}{859--864}.
\newblock


\bibitem[\protect\citeauthoryear{Simister and Brickley}{Simister and
  Brickley}{2013}]%
        {Simister2013}
\bibfield{author}{\bibinfo{person}{Shawn Simister} {and} \bibinfo{person}{Dan
  Brickley}.} \bibinfo{year}{2013}\natexlab{}.
\newblock \showarticletitle{Simple application-specific constraints for rdf
  models}. In \bibinfo{booktitle}{\emph{RDF Validation Workshop. Practical
  Assurances for Quality RDF Data, Cambridge, Ma, Boston}}.
\newblock
\urldef\tempurl%
\url{https://www.w3.org/2012/12/rdf-val/}
\showURL{%
\tempurl}


\bibitem[\protect\citeauthoryear{Simsek, Angele, K{\"{a}}rle, Opdenplatz,
  Sommer, Umbrich, and Fensel}{Simsek et~al\mbox{.}}{2021}]%
        {SimsekAKOSUF21}
\bibfield{author}{\bibinfo{person}{Umutcan Simsek}, \bibinfo{person}{Kevin
  Angele}, \bibinfo{person}{Elias K{\"{a}}rle}, \bibinfo{person}{Juliette
  Opdenplatz}, \bibinfo{person}{Dennis Sommer}, \bibinfo{person}{J{\"{u}}rgen
  Umbrich}, {and} \bibinfo{person}{Dieter Fensel}.}
  \bibinfo{year}{2021}\natexlab{}.
\newblock \showarticletitle{Knowledge Graph Lifecycle: Building and Maintaining
  Knowledge Graphs}. In \bibinfo{booktitle}{\emph{Proceedings of the 2nd
  International Workshop on Knowledge Graph Construction (KGC 2021) co-located
  with 18th Extended Semantic Web Conference (ESWC 2021), Online, June 6th,
  2021}} \emph{(\bibinfo{series}{{CEUR} Workshop Proceedings},
  Vol.~\bibinfo{volume}{2873})}. \bibinfo{publisher}{CEUR-WS.org}.
\newblock


\bibitem[\protect\citeauthoryear{Speck and Ngomo}{Speck and Ngomo}{2019}]%
        {SpeckN19}
\bibfield{author}{\bibinfo{person}{Ren{\'{e}} Speck} {and}
  \bibinfo{person}{Axel{-}Cyrille~Ngonga Ngomo}.}
  \bibinfo{year}{2019}\natexlab{}.
\newblock \showarticletitle{Leopard - {A} baseline approach to attribute
  prediction and validation for knowledge graph population}.
\newblock \bibinfo{journal}{\emph{Journal of Web Semantics}}
  \bibinfo{volume}{55} (\bibinfo{year}{2019}), \bibinfo{pages}{102--107}.
\newblock


\bibitem[\protect\citeauthoryear{Steer, Miller, and Brickley}{Steer
  et~al\mbox{.}}{2004}]%
        {TreeHugger2004}
\bibfield{author}{\bibinfo{person}{Damian Steer}, \bibinfo{person}{Libby
  Miller}, {and} \bibinfo{person}{Dan Brickley}.}
  \bibinfo{year}{2004}\natexlab{}.
\newblock \showarticletitle{Validating RDF with TreeHugger and Schematron}.
\newblock \bibinfo{journal}{\emph{w3.org}} (\bibinfo{year}{2004}).
\newblock
\urldef\tempurl%
\url{https://www.w3.org/2001/sw/Europe/events/foaf-galway/papers/pp/validating\_rdf/}
\showURL{%
\tempurl}


\bibitem[\protect\citeauthoryear{Syed, R{\"{o}}der, and Ngomo}{Syed
  et~al\mbox{.}}{2018}]%
        {SyedRN2018}
\bibfield{author}{\bibinfo{person}{Zafar~Habeeb Syed}, \bibinfo{person}{Michael
  R{\"{o}}der}, {and} \bibinfo{person}{Axel{-}Cyrille~Ngonga Ngomo}.}
  \bibinfo{year}{2018}\natexlab{}.
\newblock \showarticletitle{FactCheck: Validating {RDF} Triples Using Textual
  Evidence}. In \bibinfo{booktitle}{\emph{Proceedings of the 27th {ACM}
  International Conference on Information and Knowledge Management,
  {(CIKM2018)}, Torino, Italy, October 22-26, 2018}}.
  \bibinfo{publisher}{{ACM}}, \bibinfo{pages}{1599--1602}.
\newblock


\bibitem[\protect\citeauthoryear{Syed, R{\"{o}}der, and Ngomo}{Syed
  et~al\mbox{.}}{2019}]%
        {SyedRN2019}
\bibfield{author}{\bibinfo{person}{Zafar~Habeeb Syed}, \bibinfo{person}{Michael
  R{\"{o}}der}, {and} \bibinfo{person}{Axel{-}Cyrille~Ngonga Ngomo}.}
  \bibinfo{year}{2019}\natexlab{}.
\newblock \showarticletitle{Unsupervised Discovery of Corroborative Paths for
  Fact Validation}. In \bibinfo{booktitle}{\emph{Proceedings of the 18th
  International Semantic Web Conference {(ISWC2019)}, Auckland, New Zealand,
  October 26-30, 2019}} \emph{(\bibinfo{series}{Lecture Notes in Computer
  Science}, Vol.~\bibinfo{volume}{11778})}. \bibinfo{publisher}{Springer},
  \bibinfo{pages}{630--646}.
\newblock


\bibitem[\protect\citeauthoryear{Thorne and Vlachos}{Thorne and
  Vlachos}{2017}]%
        {ThorneV17}
\bibfield{author}{\bibinfo{person}{James Thorne} {and} \bibinfo{person}{Andreas
  Vlachos}.} \bibinfo{year}{2017}\natexlab{}.
\newblock \showarticletitle{An Extensible Framework for Verification of
  Numerical Claims}. In \bibinfo{booktitle}{\emph{Proceedings of the 15th
  Conference of the European Chapter of the Association for Computational
  Linguistics {(EACL2017)}, Valencia, Spain, April 3-7, 2017}}.
  \bibinfo{publisher}{Association for Computational Linguistics},
  \bibinfo{pages}{37--40}.
\newblock


\bibitem[\protect\citeauthoryear{Vaidyambath, Debattista, Srivatsa, and
  Brennan}{Vaidyambath et~al\mbox{.}}{2019}]%
        {VaidyambathDSB19}
\bibfield{author}{\bibinfo{person}{Ramneesh Vaidyambath},
  \bibinfo{person}{Jeremy Debattista}, \bibinfo{person}{Neha Srivatsa}, {and}
  \bibinfo{person}{Rob Brennan}.} \bibinfo{year}{2019}\natexlab{}.
\newblock \showarticletitle{An Intelligent Linked Data Quality Dashboard}. In
  \bibinfo{booktitle}{\emph{Proceedings for the 27th {AIAI} Irish Conference on
  Artificial Intelligence and Cognitive Science, Galway, Ireland, December 5-6,
  2019}} \emph{(\bibinfo{series}{{CEUR} Workshop Proceedings},
  Vol.~\bibinfo{volume}{2563})}, \bibfield{editor}{\bibinfo{person}{Edward
  Curry}, \bibinfo{person}{Mark~T. Keane}, \bibinfo{person}{Adegboyega Ojo},
  {and} \bibinfo{person}{Dhaval Salwala}} (Eds.).
  \bibinfo{publisher}{CEUR-WS.org}, \bibinfo{pages}{341--352}.
\newblock


\bibitem[\protect\citeauthoryear{Volz, Bizer, Gaedke, and Kobilarov}{Volz
  et~al\mbox{.}}{2009}]%
        {VolzBGK2009}
\bibfield{author}{\bibinfo{person}{Julius Volz}, \bibinfo{person}{Christian
  Bizer}, \bibinfo{person}{Martin Gaedke}, {and} \bibinfo{person}{Georgi
  Kobilarov}.} \bibinfo{year}{2009}\natexlab{}.
\newblock \showarticletitle{Discovering and Maintaining Links on the Web of
  Data}. In \bibinfo{booktitle}{\emph{Proceedings of the 8th International
  Semantic Web Conference (ISWC 2009), Chantilly, USA, October 25-29, 2009}}
  \emph{(\bibinfo{series}{LNCS}, Vol.~\bibinfo{volume}{5823})}.
  \bibinfo{publisher}{Springer}, \bibinfo{pages}{650--665}.
\newblock


\bibitem[\protect\citeauthoryear{Wang}{Wang}{1998}]%
        {Wang98}
\bibfield{author}{\bibinfo{person}{Richard~Y. Wang}.}
  \bibinfo{year}{1998}\natexlab{}.
\newblock \showarticletitle{A Product Perspective on Total Data Quality
  Management}.
\newblock \bibinfo{journal}{\emph{Commun. {ACM}}} \bibinfo{volume}{41},
  \bibinfo{number}{2} (\bibinfo{year}{1998}), \bibinfo{pages}{58--65}.
\newblock


\bibitem[\protect\citeauthoryear{Wang, Ziad, and Lee}{Wang
  et~al\mbox{.}}{2001}]%
        {WangZL01}
\bibfield{author}{\bibinfo{person}{Richard~Y. Wang}, \bibinfo{person}{Mostapha
  Ziad}, {and} \bibinfo{person}{Yang~W. Lee}.} \bibinfo{year}{2001}\natexlab{}.
\newblock \bibinfo{booktitle}{\emph{Data Quality}}. \bibinfo{series}{Advances
  in Database Systems}, Vol.~\bibinfo{volume}{23}.
\newblock \bibinfo{publisher}{Kluwer}.
\newblock
\showISBNx{0-7923-7215-8}


\bibitem[\protect\citeauthoryear{Wienand and Paulheim}{Wienand and
  Paulheim}{2014}]%
        {WienandP2014}
\bibfield{author}{\bibinfo{person}{Dominik Wienand} {and}
  \bibinfo{person}{Heiko Paulheim}.} \bibinfo{year}{2014}\natexlab{}.
\newblock \showarticletitle{Detecting Incorrect Numerical Data in DBpedia}. In
  \bibinfo{booktitle}{\emph{The Semantic Web: Trends and Challenges - 11th
  International Conference, {ESWC} 2014, Anissaras, Crete, Greece, May 25-29,
  2014. Proceedings}} \emph{(\bibinfo{series}{Lecture Notes in Computer
  Science}, Vol.~\bibinfo{volume}{8465})},
  \bibfield{editor}{\bibinfo{person}{Valentina Presutti},
  \bibinfo{person}{Claudia d'Amato}, \bibinfo{person}{Fabien Gandon},
  \bibinfo{person}{Mathieu d'Aquin}, \bibinfo{person}{Steffen Staab}, {and}
  \bibinfo{person}{Anna Tordai}} (Eds.). \bibinfo{publisher}{Springer},
  \bibinfo{pages}{504--518}.
\newblock
\urldef\tempurl%
\url{https://doi.org/10.1007/978-3-319-07443-6\_34}
\showDOI{\tempurl}


\bibitem[\protect\citeauthoryear{Zaveri, Rula, Maurino, Pietrobon, Lehmann, and
  Auer}{Zaveri et~al\mbox{.}}{2016}]%
        {ZaveriRMPLA16}
\bibfield{author}{\bibinfo{person}{Amrapali Zaveri}, \bibinfo{person}{Anisa
  Rula}, \bibinfo{person}{Andrea Maurino}, \bibinfo{person}{Ricardo Pietrobon},
  \bibinfo{person}{Jens Lehmann}, {and} \bibinfo{person}{S{\"{o}}ren Auer}.}
  \bibinfo{year}{2016}\natexlab{}.
\newblock \showarticletitle{Quality assessment for Linked Data: {A} Survey}.
\newblock \bibinfo{journal}{\emph{Semantic Web}} \bibinfo{volume}{7},
  \bibinfo{number}{1} (\bibinfo{year}{2016}), \bibinfo{pages}{63--93}.
\newblock


\end{thebibliography}

\end{document}